\documentclass[preprint]{aastex}

\usepackage{emulateapj5}
\usepackage{epsf}
\usepackage{psfig}
\usepackage{graphicx}
\usepackage{lscape}

\begin{document}
\title{The Implications of M Dwarf Flares on the Detection and Characterization of Exoplanets at Infrared Wavelengths}
\author{Benjamin M. Tofflemire\altaffilmark{1,2,3}, John P. Wisniewski\altaffilmark{1,3}, Adam F. Kowalski\altaffilmark{1,3}, Sarah J. Schmidt\altaffilmark{1,3}, Praveen Kundurthy\altaffilmark{1}, Eric J. Hilton\altaffilmark{1,4}, Jon A. Holtzman\altaffilmark{5}, \& Suzanne L. Hawley\altaffilmark{1}}

\altaffiltext{1}{Astronomy Department, University of Washington, Box 351580, Seattle, WA 98195, USA; tofflb@u.washington.edu, jwisnie@u.washington.edu}
\altaffiltext{2}{Current address: Astronomy Department, University of Wisconsin-Madison, 475 N Charter St, Madison, WI 53706, USA}  
\altaffiltext{3}{Visiting Astronomer, Kitt Peak National Observatory, National Optical Astronomy Observatory, which is operated by the Association of Universities for Research in Astronomy, under contract with the National Science Foundation.}
\altaffiltext{4}{Current address: Department of Geology and Geophysics, University of Hawaii, Honolulu, HI 96822, USA}
\altaffiltext{5}{Department of Astronomy, New Mexico State University, Box 30001, Las Cruces, NM 880033, USA}

\begin{abstract} 
 
 We present the results of an observational campaign which obtained high time cadence, high precision, simultaneous optical and IR photometric observations of three M dwarf flare stars for 47 hours.  The campaign was designed to characterize the behavior of energetic flare events, which routinely occur on M dwarfs, at IR wavelengths to milli-magnitude precision, and quantify to what extent such events might influence current and future efforts to detect and characterize extrasolar planets surrounding these stars.  We detected and characterized four highly energetic optical flares having \textit{U}-band total energies of $\sim$7.8 x 10$^{30}$ to $\sim$1.3 x 10$^{32}$ ergs, and found no corresponding response in the \textit{J}, \textit{H}, or \textit{Ks} bandpasses at the precision of our data.  For active dM3e stars, we find that a $\sim$1.3 x 10$^{32}$ erg  \textit{U}-band flare ($\Delta$\textit{U}$_{max}$ $\sim$1.5 mag) will induce $<$8.3 (\textit{J}), $<$8.5 (\textit{H}), and $<$11.7 (\textit{Ks}) milli-mags of a response.  A flare of this energy or greater should occur less than once per 18 hours.  For active dM4.5e stars,  we find that a $\sim$5.1 x 10$^{31}$ erg U-band flare ($\Delta$\textit{U}$_{max}$ $\sim$1.6 mag) will induce $<$7.8 (\textit{J}), $<$8.8 (\textit{H}), and $<$5.1 (\textit{Ks}) milli-mags of a response.  A flare of this energy or greater should occur less than once per 10 hours.  No evidence of stellar variability not associated with discrete flare events was observed at the level of $\sim$3.9 milli-mags over 1 hour time-scales and at the level of $\sim$5.6 milli-mags over 7.5 hour time-scales.  We therefore demonstrate that most M dwarf stellar activity and flares will not influence IR detection and characterization studies of M dwarf exoplanets above the level of $\sim$5-11 milli-mags, depending on the filter and spectral type.  We speculate that the most energetic megaflares on M dwarfs, which occur at rates of once per month, are likely to be easily detected in IR observations with sensitivity of tens of milli-mags.  We also discuss how recent detections of line flux enhancements during M dwarf flares could influence IR transmission spectroscopic observations of M dwarf exoplanets.

\end{abstract}


\section{Introduction} \label{intro}

Pioneering exoplanet studies using the excellent sensitivity and stability of the Spitzer Space Telescope have begun to characterize some of the basic properties of exoplanets (see e.g. \citealt{sea10}).  Secondary transit observations have enabled determination of the temperatures of exoplanets \citep{cha05,ago10}, while monitoring of infrared (IR) flux as a function of orbital phase, i.e. ``phase-mapping'', has been used to place constraints on the atmospheric recirculation efficiency of exoplanetary atmospheres \citep{knu07,knu09}.  With respect to exoplanet detections, re-evaluations of the impact of stellar activity and flares from M dwarfs on the habitability of planets they harbor indicate that these systems could be amenable to hosting habitable exoplanets \citep{tar07,seg10}.  Super-Earth size exoplanets in the habitable zone of M dwarfs can be detected via the transit technique using existing instrumentation; hence, survey programs such as MEarth are targeting large numbers of M dwarfs to find these types of planets \citep{irw09}.

Low-mass stars are well known to exhibit a range of stellar activity phenomena \citep{haw93}, and the prevalence of active stars is observed to decrease as a function of age (see e.g. \citealt{wes08}).  One of the more energetic byproducts of stellar activity are flares, which are thought to be produced by magnetic reconnection in the atmospheres of cool stars.  Active M dwarfs exhibit some of the more energetic and highest rate of flares amongst cool stars \citep{lac76,haw91,kow09,kow10}.  M dwarf flares can last from minutes to many hours, exhibit U-band flux enhancements of a factor of several hundred above quiescence, and emit energies of up to E$_{U,flare}$ $>$ 10$^{34}$ ergs \citep{haw91,cul94,kow10,ost10}.  The optical white-light emission has been shown to be fit by a two component model comprised of Balmer continuum emission plus a T $\sim$10,000K blackbody component \citep{kow10}.  Although M dwarf flares are most conspicuous at radio, blue optical, and X-ray wavelengths \citep{haw03,ost05}, \citet{sch11} has recently presented the clear detection of IR (0.95-2.46$\mu$m) flare signatures in H I Paschen lines, H I Br$\gamma$, and the He I 10830 \AA\ line during events confirmed with simultaneous optical photometry.  Several solar flares have also been detected in the continuum region near 1.56$\mu$m in the IR, near the H$^{-}$ opacity minimum, implying formation deep within the solar atmosphere \citep{xu06,che10}.

Some types of stellar activity have already been observed to ``contaminate'' exoplanet detection and characterization studies.  Variations in the appearance and depth of transit light curves have been observed for FGKM host stars \citep{ben09,dit09,ago10,kun11} and in transmission spectra of the M dwarf Gl 436 \citep{knu11}.  Moreover, \citet{knu10} suggest that UV flux from chromospherically active stars may destroy the atmospheric compounds responsible for producing temperature inversions observed in the atmospheres of exoplanets orbiting less active stars.  

Although M dwarfs are known to exhibit a range of activity, the magnitude of such variability in IR continuum bandpasses relevant to the detection (via primary transits) and characterization (via secondary transits and phase mapping) of exoplanets remains poorly understood.  Marginal variability that may be due to flares was seen in a statistical investigation of single-epoch 2MASS calibration data \citep{dav11}.  That study is complementary to the simultaneous monitoring observations reported here, which provide
flare rates for individual stars.  The few previous simultaneous optical-IR observations of M dwarf flares report anomalous IR flux decrements during optical flares \citep{rod88} or null detections \citep{pan95}.  We caution however that the observational techniques and analyses of these simultaneous photometric studies were dubious and not optimized for precision IR photometry.  

Could the 5 milli-mag transit signal of a super-Earth residing in the habitable zone around a M5V star \citep{irw09} be routinely masked by stellar activity and flares when observed in the IR?  In this paper, we present simultaneously obtained, high time cadence optical and IR photometric monitoring of well known active M dwarfs at milli-mag precision to address this question.

\section{Observations and Data Reduction} \label{obs-red}

We used a suite of 3 optical photometric facilities and 1 IR photometric facility to obtain simultaneous broad-band optical and IR photometry of three active M dwarfs known to exhibit frequent flares, between 2009 October 2 - 7 and 2011 February 11 - 16.  The blue optical photometric data were obtained to identify the onset and characterize the energy of flare events, whereas the red optical photometry was used to help characterize the color of flares.  A basic summary of the targets observed at each observational facility, along with exposure times, filters used, and observing cadences is presented in Table \ref{obssum}.  Most of our 2009 observing run was lost due to a tropical storm hitting KPNO, while poor weather and instrument failures only affected $\sim$2 nights of our 2011 observing run.

\subsection{Optical Photometry}

The Astrophysical Research Consortium Small Aperture Telescope (ARCSAT) 0.5m at the Apache Point Observatory (APO) was one of three facilities we used to obtain high candence optical photometry.  Observations were made with the University of Washington \textit{Flare-Cam} \citep{hi11a,hil11} through Sloan \textit{g,r,i} filters, and were recorded using a thermo-electic cooled (TEC) 1024 x 1024 pixel APOGEE CCD.  The full frame readout of the 8$\farcm$ x 8$\farcm$ FOV was $\sim$1 second.  Twilight sky flats and dark frames were obtained every night to aid reduction of the data.  The data were reduced using standard techniques, and differential aperture photometry was extracted using a Python-based script which called the IRAF \textit{phot} package.  

The New Mexico State University (NMSU) automated 1m telescope at APO also provided high cadence Johnson \textit{U} band observations.  Further details about the NMSU 1m and its operations are described by \citet{hol10}.  The data were reduced using an automatic photometry pipeline, yielding differential photometry for our science targets.

During our 2009 observing run we also obtained simultaneous optical monitoring of our science targets with the Wisconsin Indiana Yale NOAO (WIYN) 0.9m telescope.  We used the S2KB CCD, a 2048 x 2048 pixel camera with a pixel scale of 0$\farcs$60 pixel$^{-1}$.  We employed 2 x 2 binning of the S2KB chip, and a 260 x 260 pixel sub-array of the full frame, to reduce read out time and obtain higher cadence photometry.  The data we present were obtained in the Harris \textit{U} band filter.  Nightly bias and dome flat field frames were obtained, and the data were reduced using standard IRAF techniques.  Differential photometry was extracted using the same Python-based software implemented for our ARCSAT data.

\subsection{IR Photometry} \label{obs-ir}

Our IR photometry was obtained at the KPNO 2.1m telescope using the Simultaneous Quad Infrared Imaging Device (SQIID).  SQIID obtains simultaneous \textit{J} ($\lambda_{center}$ = 1.267 $\mu$m), \textit{H} ($\lambda_{center}$ = 1.672 $\mu$m), and \textit{Ks} ($\lambda_{center}$ = 2.224 $\mu$m) band imagery in 3 of the 4 quadrants of its array.  At the f/15 focus, each 512 x 512 quadrant of SQIID's ALADDIN array spans a field of view of 5$\farcm$07 x 5$\farcm$28 with a pixel scale of 0.69 arcsec pixel$^{-1}$.  Our science exposures were generally obtained at or near the array's minimum exposure time of 0.874 seconds, and we defocussed the telescope until stars took the appearance of $\sim$20$\farcs$ - $\sim$28$\farcs$ -wide donuts, to keep the counts below the array's linearity limit.  Because of large overheads in the transmission of data from the array, our net observing cadence was $\sim$60 seconds (Table \ref{obssum}).

Night-time, deep blank-sky images were obtained every clear night to serve as sky flats; dark frames were also obtained every day corresponding to the integration times of our sky flats.  We used a 3-4 dither point pattern in order to mitigate ghosts produced from observing bright sources.  Since we were interested in performing differential photometry, the specific number and pattern of dither points we adopted varied by source, to ensure each frame contained our science target and at least 1 comparison star.  We periodically adjusted the position of the science target on the array to ensure that small amplitude, long time-scale telescope tracking imperfections did not cause the target to fall on different pixels, thus making it more susceptible to imperfections in the flat fielding process.

We developed a custom data reduction routine to extract the highest precision photometry possible.  We first applied a third order linearity correction to the data using coefficients in the SQIID user's manual, and then divided these data by a normalized flat field image. Sky subtraction and dark correction of our science images was achieved by subtracting immediately adjacent dither frames from each other.  Individual pixels $>$16,000 ADUs (the linearity limit) or $<$ -500 ADUs were next flagged as bad or warm pixels, and corrected using the IRAF routine \textit{fixpix}.  Next, low-level array artifacts, such as column banding and less frequently observed row banding, were characterized by sampling this structure in star-free regions of the array and removed via custom IDL routines.  

To obtain photometry for our SQIID data, we first identified the centroid position of each star in our FOVs using SExtractor \citep{ber96}, and a custom reference profile which mimicked the extended donut-shape of our sources.  Next, we ran these stellar positions through IRAF's \textit{phot} routine to extract aperture photometry, using a target aperture of radius 20 pixels (EV Lac and YZ CMi) to 24 pixels (AD Leo) and a 4 pixel wide background sky annulus starting at a radius of 30 pixels from each centroid position.  We then extracted differential photometry for our science targets using the brightest, and often only, comparison star with the SQIID FOV, as summarized in Table \ref{comp}.  Previous IR monitoring programs dedicated to observing M dwarfs for transits (MEarth; \citealt{irw11}) or high precision observations of brown dwarfs to characterize photospheric condensates \citep{bai03} have described how variations in relative humidity can affect IR differential photometry.  We found we were able to remove most of the small amplitude, long time-scale photometric trends in our data likely attributable to these effects by simply fitting a linear function to each dither position's light curve individually.  Finally, each dither position was normalized to a uniform scale and combined to yield the highest time cadence differential photometry from our observations.

Table \ref{flaresum} summarizes the relative photometric stability we were able to obtain for every target and filter, both across an entire night and across 1-hour windows, quantified as the observed standard deviation.  The IR stability we achieved for AD Leo, 9.7-12.6 milli-mags over a full night and 8.4-10.5 milli-mags over a 1 hour window, was systematically larger than that achieved for our other targets due to the lower flux of the brightest comparison star in its field of view.  The best IR photometric stability we achieved was an impressive 5.1-6.0 milli-mags over a full night and 3.8-3.9 milli-mags over a 1 hour window for our 2011 February 13 monitoring of YZ CMi.

\section{Results} \label{results}

\begin{figure}
\begin{center}
\includegraphics[width=14cm,angle=90]{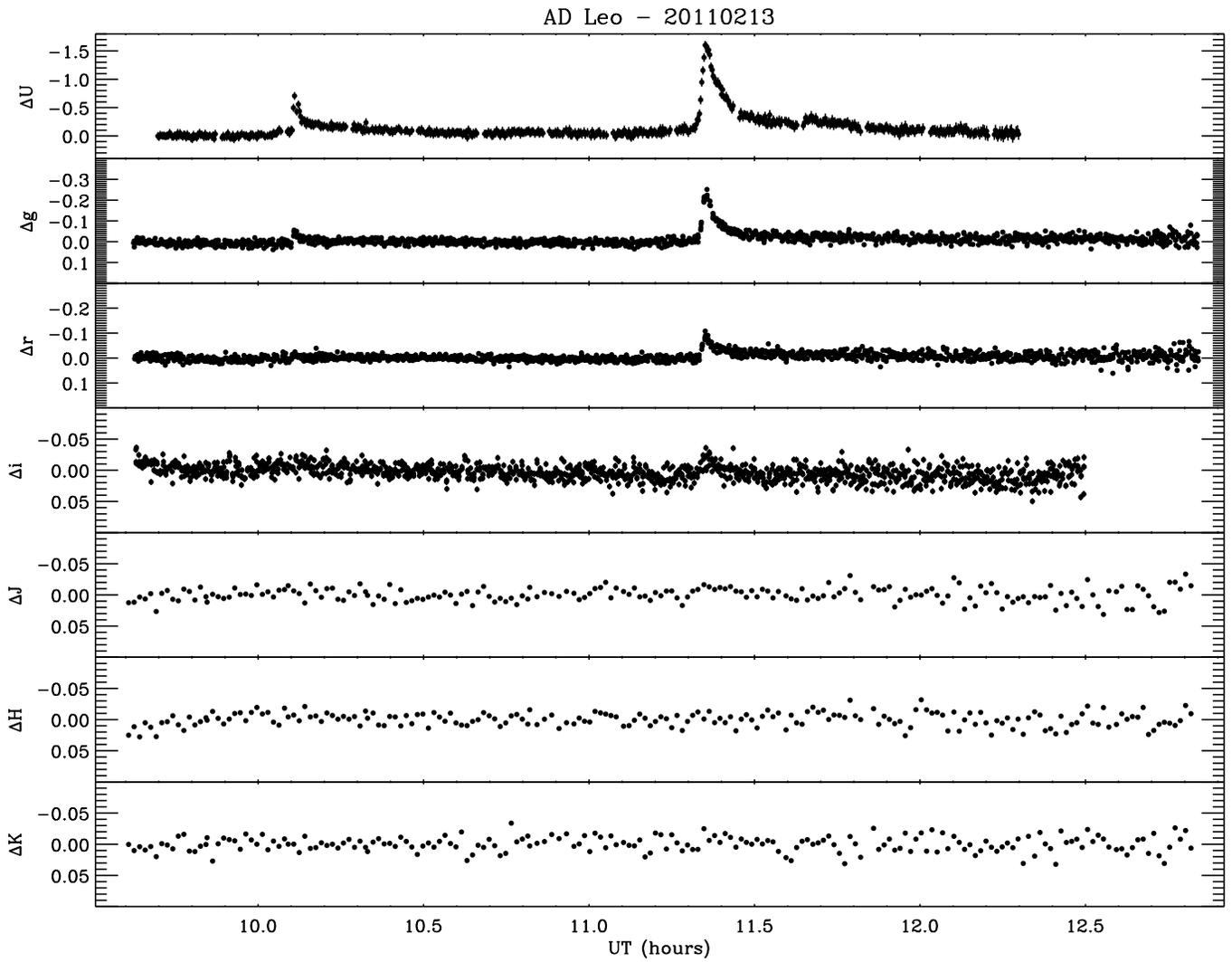}
\caption{The differential photometric light curve for our observations of AD Leo on 
the night of 2011 February 13 is shown for the U, g, r, i, J, H, and Ks filters.  Photon statistics-based error bars are plotted in all panels, but these error bars are generally smaller than the size of the data points  \label{adleoall}}
\end{center}
\end{figure}

Figure \ref{adleoall}  shows a representative multi-filter differential photometric light curve obtained for the star AD Leo on 2011 February 13.  Note that photon statistics-based error bars are plotted in all panels of Figure \ref{adleoall}, and for all figures in this paper, but these error bars are generally smaller than the size of the data points.  Before characterizing the behavior of stellar flares in our IR data, we first identified these events in our simultaneously obtained optical photometry.  We used the IDL-based flare finding software described in \citet{hi11a} and \citet{hil11}.  Flares are identified as a single epoch which is 3.5-$\sigma$ brighter than the standard deviation of the local mean, followed by three epochs which are 2-$\sigma$ above the local mean.  Each flare identified was also double-checked by eye to confirm the event was robust.  The total number of flares detected in each bandpass using this flare finding algorithm is shown in Table \ref{flaresum}; a total of 20 flares were detected in our \textit{U}-band data.  

For the larger optical flare events observed, additional flare properties such as the peak magnitude enhancement and total flare duration are compiled in Table \ref{flarenergy}.  Flare equivalent durations, defined as the amount of time each star would need to spend at a quiescent level to produce the same total energy as during each flare \citep{ger72}, were determined via our flare finding software by integration under each light curve.    We adopted the quiescent luminosities for AD Leo (1.59 x 10$^{29}$ ergs s$^{-1}$), EV Lac (6.63 x 10$^{28}$ ergs s$^{-1}$), and YZ CMi (4.57 x 10$^{28}$ ergs s$^{-1}$) computed by \citet{hi11a} and \citet{hil11}, and based on data in \citet{rei95}.  We then multiplied the quiescent luminosities by the equivalent duration of each event, to extract flare energies (Table \ref{flarenergy}).

\subsection{Optical Flares} \label{opticalflares}

\begin{figure}
\begin{center}
\includegraphics[width=14cm,angle=90]{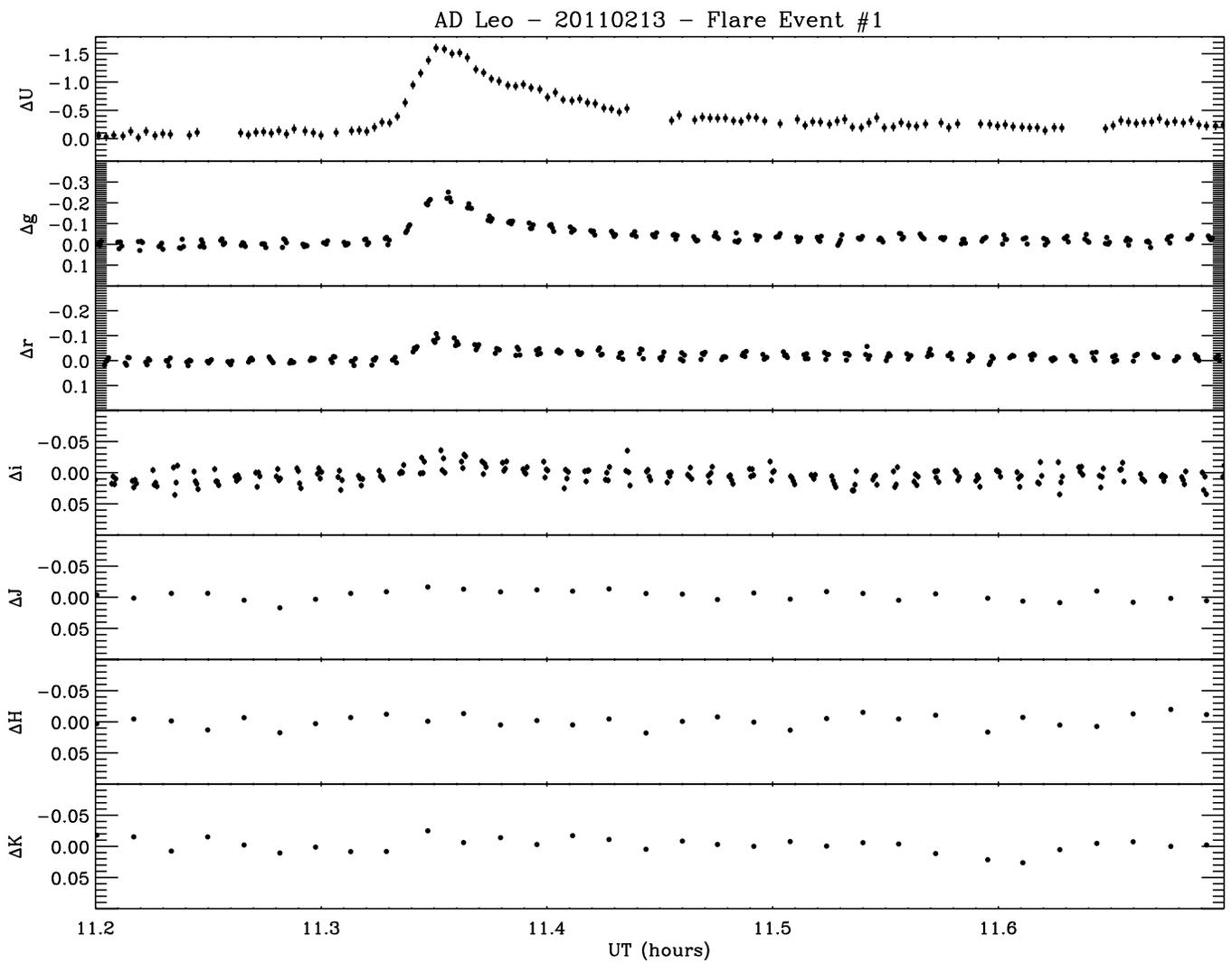}
\caption{A zoomed view of the large flare observed on AD Leo on 2011 February 13 (Figure \ref{adleoall}), referred to as flare event \#1 in Table \ref{flarenergy}.  The online version of this journal contains analogous figures for flare events \#2 - \#4. 
 \label{adleozoom}}
\end{center}
\end{figure}

As the focus of this paper is on characterizing the properties of stellar flares in the IR, we focus our attention on the four largest U-band flares 
we observed on AD Leo (event \#1), YZ CMi (events \#2 \& \#3), and EV Lac (event \#4), as listed in Table \ref{flarenergy} and illustrated in Figure \ref{adleozoom}.  Our flares exhibited \textit{U}-band energies ranging from $\sim$7.8 x 10$^{30}$ (event \#4) to $\sim$1.3 x 10$^{32}$ (event \#1) ergs.  Flare events \#1-3 are considered strong \citep{lac76} although they are less energetic than the very rare $\sim$10$^{34}$ erg megaflare \citet{kow10} observed on YZ CMi. 

For each of our targets, \citet{lac76} presented flare frequency distributions which follow power laws, except at the lowest and highest energy regimes.  \citet{hi11a} and \citet{hil11} expanded upon this earlier work, presenting flare frequency distributions for a wider range of M dwarfs.  After correcting for detection efficiency, \citet{hi11a} was able to demonstrate that flare frequency distributions follow power law distributions even down to low energies.  We use the flare frequency distributions presented in \citet{hi11a} and \citet{hil11} for this paper.  Note that since the flare energies we computed in Table \ref{flarenergy} used the same quiescent stellar luminosities employed to construct the Hilton et al flare frequency distributions, any future refinements of these quiescent luminosities will not influence the flare rates we hereafter cite.  

We determine that a flare having equal or greater \textit{U}-band energy to our event \#1, on AD Leo, would occur less than once per 18 hours.  For YZ CMi, we find that a flare having equal or greater \textit{U}-band energy for events \#2-3 would occur less than once per 10 hours, and the EV Lac flare frequency distribution indicates that a flare having equal or greater \textit{U}-band energy to our event \#4 would occur less than once per 7 hours.  

\subsection{IR Continuum Flares} \label{irflares}

We utilized the same flare finding algorithm described in Section \ref{results} and used to identify flares in our optical data to search for contemporaneous flares in our full set of IR data.  We detected no statistically significant deviations in our IR photometry at the epochs of the four major, optically detected 
flares.  We quantify the 1-$\sigma$ upper limits of these four events by computing the standard deviation of all data in 
a 40 minute time window before each event, and summarize these limits in Table \ref{flarenergy}.  

\begin{figure}
\begin{center}
\includegraphics[width=14cm,angle=90]{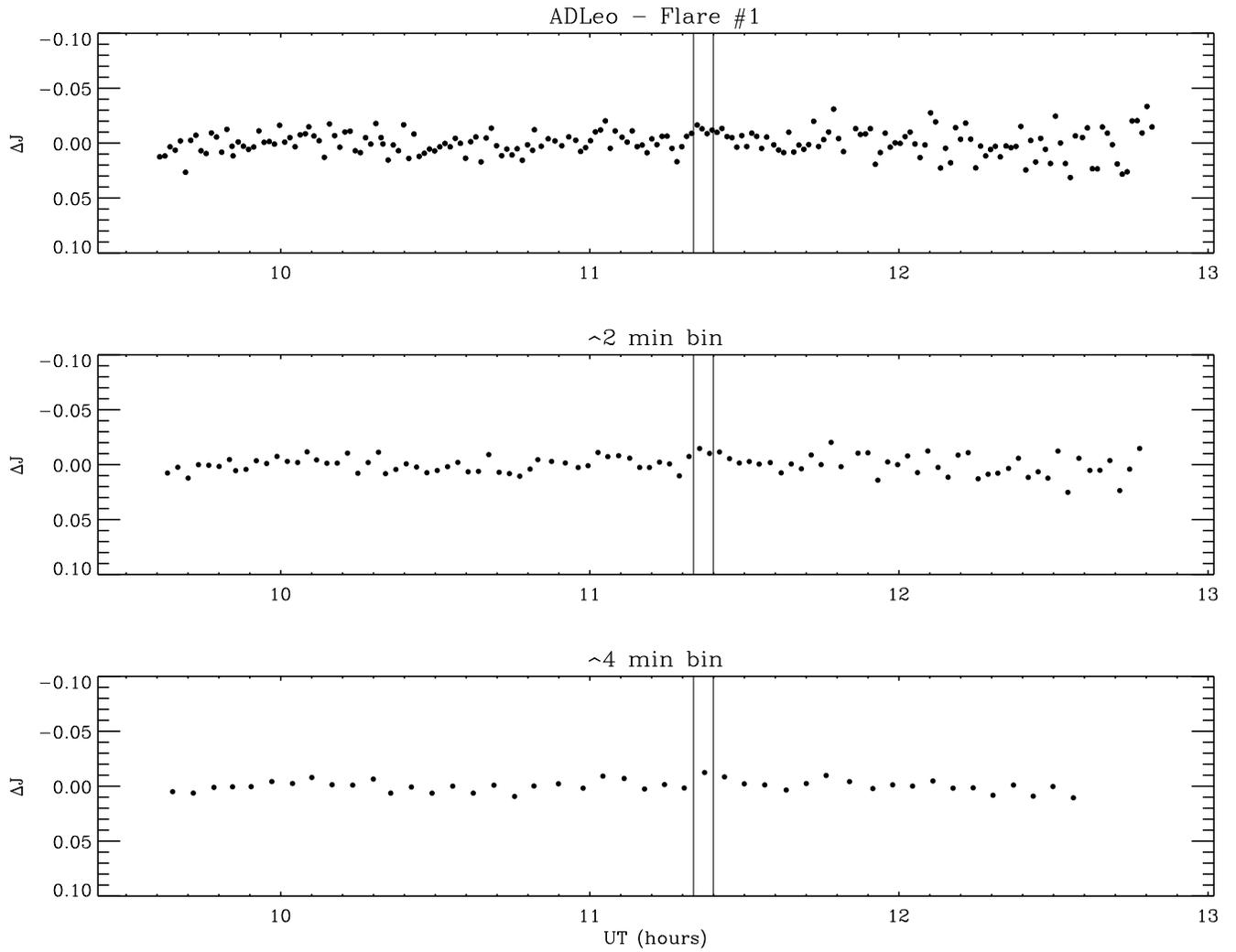}
\caption{The J-band differential photometry of flare event \#1 (Table \ref{flarenergy}), observed on AD Leo on 2011 February 13.  The $\sim$4 minute duration of the flare in the i filter, the closest filter to the J-band, is depicted by solid 
vertical lines.  The top panel depicts the native time resolution of the data ($\sim$1 minute), the middle panel depicts the data binned to a cadence of $\sim$2 minutes, and the bottom panel depicts the data binned to a cadence of $\sim$4 minutes.  The online version of this journal contains analogous J-band figures for flare events \#2 - \#4 described in Table \ref{flarenergy}.
 \label{event1J}}
\end{center}
\end{figure}

\begin{figure}
\begin{center}
\includegraphics[width=14cm,angle=90]{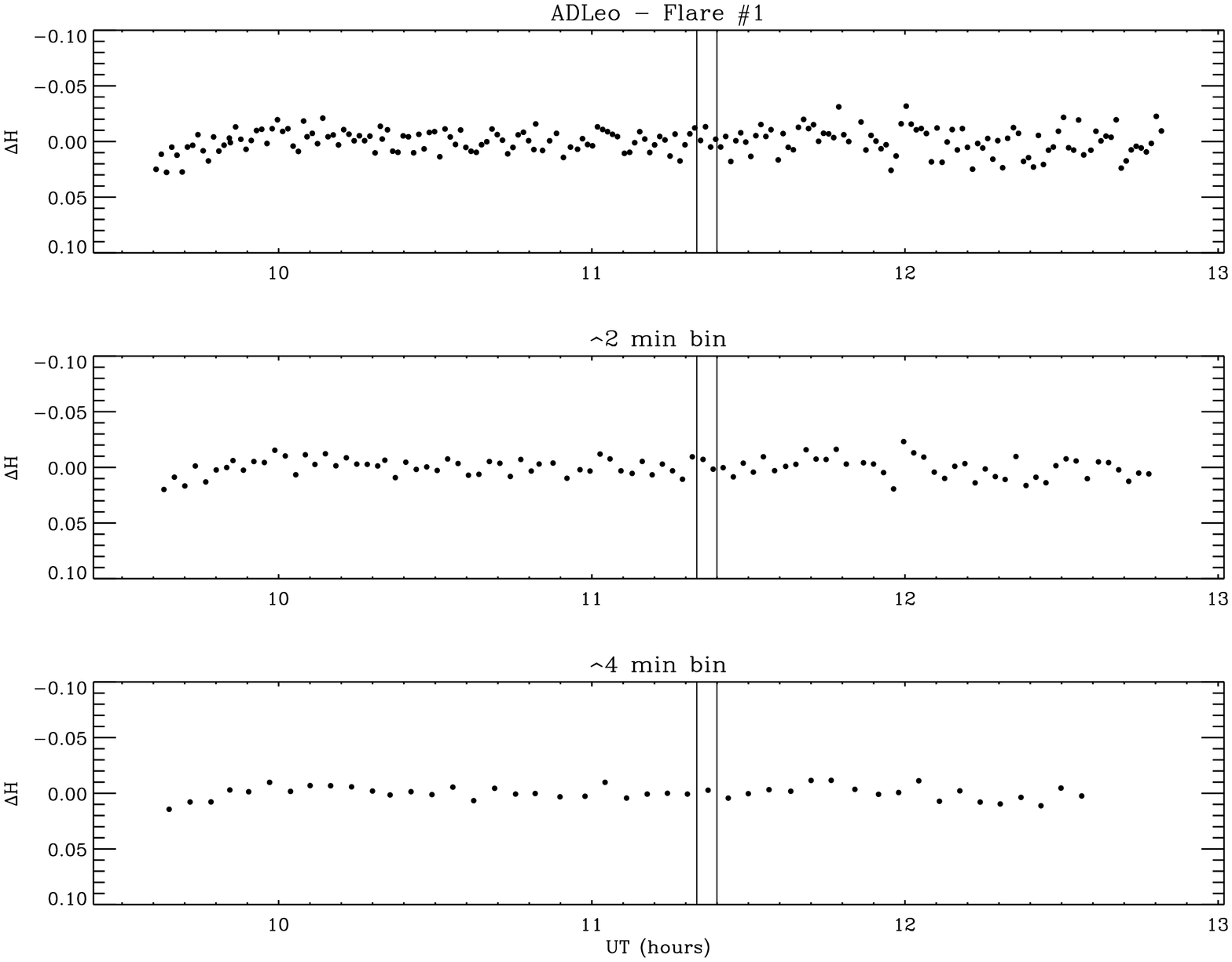}
\caption{The same as Figure \ref{event1J}, plotted instead in the H-band.  The online version of this journal contains analogous H-band figures for flare events \#2 - \#4 described in Table \ref{flarenergy}.
 \label{event1H}}
\end{center}
\end{figure}

\begin{figure}
\begin{center}
\includegraphics[width=14cm,angle=90]{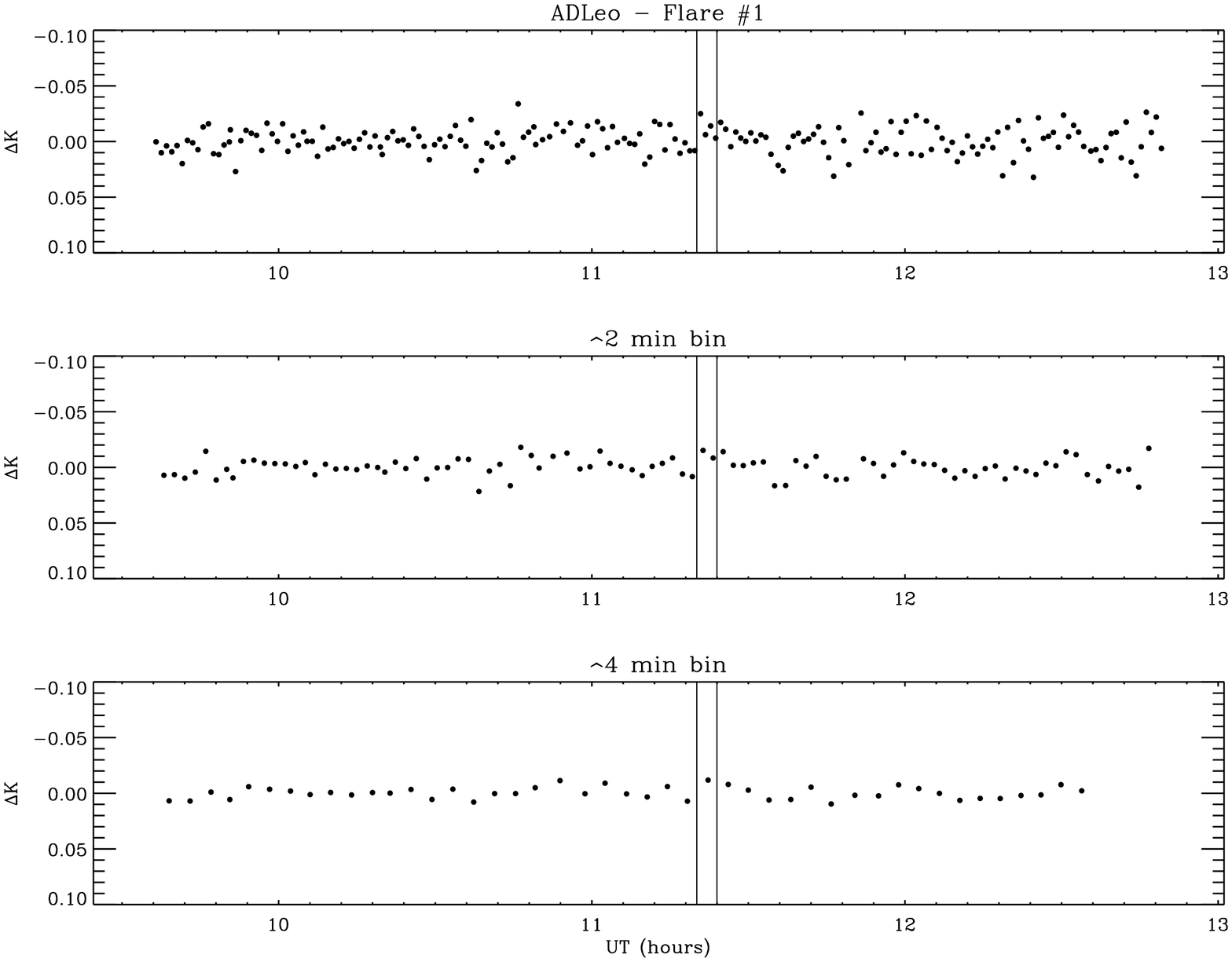}
\caption{The same as Figure \ref{event1J}, plotted instead in the Ks-band.  The online version of this journal contains analogous Ks-band figures for flare events \#2 - \#4 described in Table \ref{flarenergy}.
 \label{event1K}}
\end{center}
\end{figure}

Our \textit{U}-band photometry provides us with information about the starting time of each flare event.  As seen in Figure \ref{adleoall} and demonstrated by many others (see e.g. \citealt{haw91,hil11}), the total duration of flares in the optical is color dependent.  Flare \#1 in our survey, for example, was visible in the \textit{U} filter for 10x longer than it was in the \textit{i} filter (Table \ref{flarenergy}).  We therefore estimated upper limits to the duration which each flare event would be visible in the \textit{J}, \textit{H}, and \textit{Ks} filters by adopting the duration of each event in the reddest optical filter in which it was detected (e.g. the \textit{i}-band for event \#1).  Our observing 
cadence (Table \ref{obssum}) constrains the uncertainty in the durations we quote.  This information enabled us to time bin our data, as shown in Figures \ref{event1J}, \ref{event1H}, and \ref{event1K} for the \textit{J}, \textit{H}, and \textit{Ks} filters respectively, to intervals which would create two data points (i.e. $\sim$2 minute binning) and 1 data point (i.e. $\sim$4 minute binning) across the upper limit duration of each event.  As flare light curves can exhibit significant changes over 2-4 minute intervals (see e.g. Figure \ref{adleoall}), binning our data to these longer cadences could cause flare signatures to be diminished.  Conversely, binning will help to reduce the dispersion of quiescent-phase data, and therefore could help elucidate the presence of low-amplitude slowly varying events, such as the long-duration flare shape of \citet{haw95}.  However, even the binned data did not reveal a statistically significant event in the IR.

\begin{figure}
\begin{center}
\includegraphics[width=14cm,angle=90]{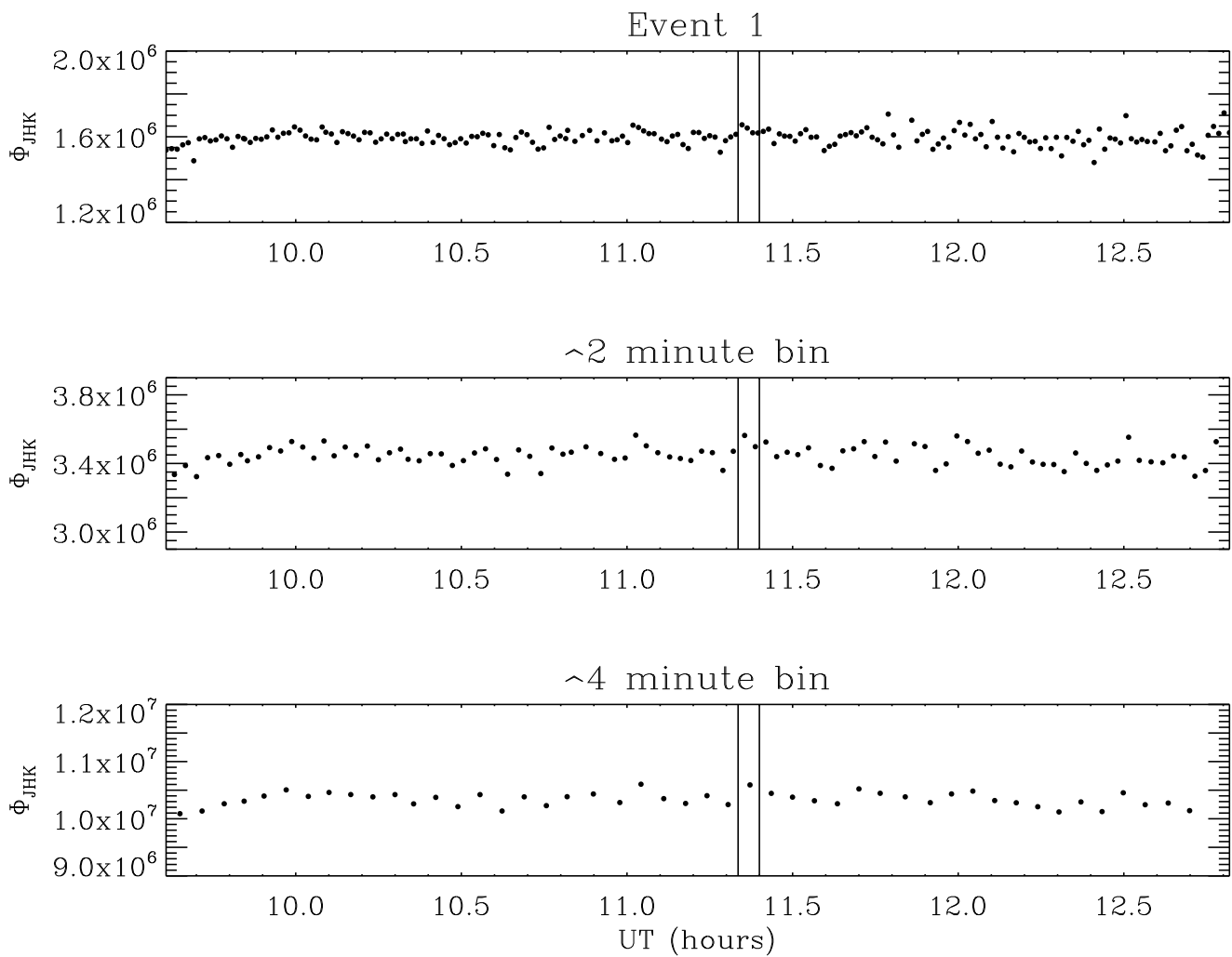}
\caption{The $\Phi_{JHK}$ statistic for flare event \#1, depicted by vertical lines, is shown.  We find no statistical evidence for an enhanced signal in this statistic for our unbinned data (top panel), our data binned in $\sim$2 minute intervals (middle panel), or in our data binned in $\sim$4 minute 
intervals (bottom panel). \label{phi1}}
\end{center}
\end{figure}

Following practices described in \citet{kow09}, we also computed a Welch and Stetson variability index \citep{wel93}, $\Phi_{JHK}$, for each epoch of our data.  This type of variability index builds on the principle that the photometric variations we are searching for should appear at similar time epochs across multiple filters; hence, it serves as a way to identify faint signals common to each of our \textit{J}, \textit{H}, and \textit{Ks} filter data.  In practice, the index is computed simply by: 

\begin{equation}
\Phi_{JHK} = \left( \frac{Flux_{J}} {\sigma_{J}}\right)   \left(\frac{Flux_{H}} {\sigma_{H}}\right)  \left(\frac{Flux_{Ks}} {\sigma_{Ks}}\right)
\end{equation}

where the flux in each multiplicative term is determined from each filter's differential photometry light curve.  The resultant $\Phi_{JHK}$ index for our unbinned, $\sim$2 minute binned, and $\sim$4 minute binned data exhibited no statistically significant change during the time of each flare event, as shown in Figure \ref{phi1} for event \#1, as compared to the value of the index preceeding and following each event.  

We can not exclude the possibility that our IR observing cadence (Table \ref{obssum}) of one (0.874 second) integration every $\sim$60 seconds does not influence our ability to detect the optically observed flare events.  To help better quantify whether our observations could miss IR responses to flares which begin and end on time-scales faster than the 
relative observing cadence we achieved, we identified the time of the peak flare emission during each event, as listed in Table \ref{flaretime}.  The quoted error bars on these times correspond to the epoch of the prior and subsequent integration in each filter.  We also list the epochs of the nearest 1-2 IR observations in Table \ref{flaretime} relative to the observed epoch of the peak $\Delta$\textit{U}-band flux.  The epochs 
of our IR observations essentially overlap with the epochs of the observed peak $\Delta$\textit{U}-band flux for flare events \#1,3,and 4 to within the timing uncertainties.  Our 
closest IR integration for flare event \#2 was 18.5 seconds after the oberved $\Delta$U-band peak, which is larger than the $\pm$13 second uncertainty in the epoch of the peak $\Delta$\textit{U}-band flux.  These data quantify the limits at which our different optical versus IR observing cadences could have resulted in the non-detection of flare events.

We conclude that, at the limit of our observational data, we find no compelling evidence of flare enhancements in our \textit{J}, \textit{H}, or \textit{Ks} filter data during the $\sim$7.8 x 10$^{30}$ to $\sim$1.3 x 10$^{32}$ erg \textit{U}-band flares we detected.

\section{Discussion}

We have presented the results of $\sim$47 hours of high cadence, high precision, simultaneous optical and IR photometric monitoring of 3 active M dwarfs.  We now discuss the interpretation of these results in the context of the effects that stellar flares could have on a variety of M dwarf exoplanet studies.  For reference, we remind the reader that a Super-Earth (2 R$_{\earth}$) residing in the habitable zone around a dM5 star ($\sim$0.074 AU separation; see e.g. \citealt{nut08}) would have an orbital period of 14.5 days and a transit duration of $\sim$2 hours.

\subsection{Effects of Stellar Flares on IR Transit and Phase Mapping of M Dwarf Exoplanets: Observational Constraints} \label{cont}

In Section \ref{irflares} we demonstrated that during four significant optical flare events, having \textit{U}-band energies ranging from $\sim$7.8 x 10$^{30}$ to $\sim$1.3 x 10$^{32}$ ergs, we observed no statistically significant evidence of corresponding broad-band enhancements in the \textit{J}, \textit{H}, and \textit{Ks} filters at the 5.1-11.7 milli-mag level.  These upper limits were computed from the standard deviation of 
all data in a 40 minute time window before each flare.  

We quantified the relative frequency that one would expect for flares with energies similar to events \#1-4 in Section \ref{opticalflares}, using flare frequency distributions observed for our specific dMe stars in \citet{hi11a} and \citet{hil11}.  By combining these two observational properties, we can place upper limits on the effects of stellar flares on future continuum-based observations of M dwarf exoplanetary systems.  For active M3Ve stars, we find that a $\sim$1.3 x 10$^{32}$ erg \textit{U}-band flare will induce $<$8.3, $<$8.5, and $<$11.7 milli-mags of an effect in the \textit{J}, \textit{H}, and \textit{Ks} filters respectively.  A flare of this energy or greater should occur less than once per 18 hours.  For active M4.5e stars,  we find that a $\sim$5.1 x 10$^{31}$ erg \textit{U}-band flare will induce $<$7.8, $<$8.8, and $<$5.1 milli-mags of an effect in the J, H, and Ks filters respectively.  A flare of this energy or greater should occur less than once per 10 hours.  Moreover, we observe no evidence of stellar variability not associated with discrete flare events at the level of $>$3.9 (\textit{J}), $>$3.8 (\textit{H}), and $>$3.9 (\textit{Ks}) milli-mags over 1 hour time-scales and no level of stellar variability at the level of $>$6.0 (\textit{J}), $>$5.6 (\textit{H}), and $>$5.1 milli-mags over 7.5 hour time-scales.  These data provide quantitative upper limits to the level of broad-band stellar variability which could be expected in IR transit and phase mapping observations of M dwarfs. 

\subsection{Effects of Stellar Flares on IR Transit and Phase Mapping of M Dwarf Exoplanets: Theoretical Constraints}

Although multi-wavelength radiative hydrodynamic modeling of solar and dMe flares has been constructed \citep{all05,all06}, we are not aware of any suite of detailed low mass star flare models in the existing literature that present model IR fluxes.  \citet{dav11} begins to mitigate this deficiency by adopting the two component flare emission model of \citet{kow10}, comprised of Balmer continuum predictions by radiative hydrodynamic flare models of \citep{all06} and a T $\sim$10,000K blackbody, and extrapolating this model to predict expected fluxes over a range of optical and IR filters.  For our flare event \#1 (Table \ref{flarenergy}), this model predicts a peak \textit{i}-band flux enhancement ($\sim$0.02 mag) which is within a factor of 2 of our observed value of 0.04 magnitudes and predicts an IR response of 3.5 milli-mags (\textit{J}-band), 2.1 milli-mags (\textit{H}-band), and 1.8 milli-mags (\textit{Ks}-band) which is consistent with the upper limits we set observationally in Section \ref{irflares}.  

Using the \citet{dav11} extrapolations, we explore the level of flare which would need to occur to be detected by our observing program, 
given the typical standard deviation of our IR data during epochs of observed flares, $\sim$8 milli-mags.   For a dM3e star, these models suggest a flare exhibiting a peak \textit{U}-band enhancement of 2.6 magnitudes (i.e. one magnitude greater than our largest flare) would have produced \textit{J}-band (11.6 milli-mags), \textit{H}-band (6.8 milli-mags), and \textit{Ks}-band (5.7 milli-mags) responses which would have been detected by our observational program.  Constraining the frequency of such large flares is difficult as observed flare frequency distributions deviate from simple power laws at high energies \citep{lac76,hi11a,hil11}.  Nonetheless, we can conservatively state that the frequency of such events should be less than once per 35 hours.

We also note that the E$_{U}$ $>$10$^{34}$ ergs megaflare on YZ CMi reported by \citet{kow10} was also detected by the MEarth monitoring 
program \citep{irw11}, which observes with an i+z filter \citep{nut08}.  This $\sim$6 magnitude U band event \citep{kow10} produced a $\sim$0.6 magnitude enhancement in the MEarth (i+z) filter (Irwin 2011; personal communication), which is generally consistent with the predicted level of enhancement from \citet{dav11}.  For that flare, the model predicts \textit{J}-band (119 milli-mag), \textit{H}-band (77 milli-mag), and \textit{Ks}-band (62 milli-mag) enhancements, which are significantly larger than our photometric noise floor.  Although we did not observe such a large flare, which \citet{kow10} suggest should occur at a rate of once per month, we speculate that such events will easily be detectable in future IR observations of M dwarfs.

The \citet{kow10} observational and modeling work which formed the basis of the extrapolations computed by \citet{dav11} is unique in that it provided convincing evidence that the blue optical properties of flares could be reproduced by 2 components, a Balmer continuum predicted by radiative hydrodynamic flare models \citep{all06} and a T $\sim$10,000K blackbody.  This framework represents an important step towards advancing our understanding of flares; however, it is phenomenological in nature and by design doesn't reproduce the entire myriad of observational characteristics of flares.  \citet{kow11}, for example, suggest that additional components from H I Paschen continuum emission and photospheric backwarming can reproduce the residual observed flux from 5000-5500 \AA, which isn't accounted for in the 2-component computation.  It is therefore plausible that including additional components relevant to the \textit{J,H,Ks} wavelength regime, such as Brackett and Pfund continuum emission, could also be important.  Although extrapolating a blue/near-UV continuum model may not yield a complete description of the IR flare continuum, we suggest extrapolations like those presented in \citet{dav11} and adopted here can provide an order of magnitude guidance on the expected IR response of flares.  Although beyond the scope of this paper, future 
models of stellar flares should consider providing broad-band flux predictions at IR wavelengths to the community, to complement 
the observational constraints provided by this work.

\subsection{Effects of Stellar Flares on Transmission Spectroscopy Studies of M Dwarf Exoplanets: Observational Constraints} \label{lines}

Since the discovery of sodium in the atmosphere of HD 209458b by \citet{cha02}, transmission spectroscopy has been recognized as one of the best ways to directly probe the chemistry of exoplanetary atmospheres.  The technique has been successfully employed with current ground- and space-based facilities, leading to detections of atmospheric molecules like water, methane, carbon monoxide, and carbon dioxide \citep{sw09a,sw09b,swa10,knu11}, and is anticipated to play an important role in characterizing the atmospheres of exoplanets once JWST is operational (see e.g. \citealt{cla10,bel11}).  In this context, we note that \citet{sch11} reported the first detection of IR line emission in H I
 Br$\gamma$ during a $\sim$4 x 10$^{32}$ erg (\textit{u}-band energy) flare on EV Lac, whereby the line flux was enhanced during the flare by $\sim$5\%
  above quiescent values.  \citet{sch11} also reported the presence of H I Pa$\beta$, Pa$\gamma$, and Pa$\delta$ emission during three separate
   flare events with \textit{u}-band (and also \textit{U}-band) energies $\sim$4 x 10$^{31}$ to $\sim$4 x 10$^{32}$ ergs, with observed line flux enhancements of 5-20\% above that
    observed in quiescence.  The expected rate, i.e. duty cycle, of IR line emission was estimated to be $\sim$3\% for active mid-M dwarfs by \citet{sch11}.  The widths of these lines are much smaller than the widths of molecular features being characterized in current exoplanet studies, and therefore should not preclude investigations of the atmospheric molecular content of exoplanets surrounding M dwarfs.  However, efforts to interpret future transmission spectroscopy observations of M dwarfs should allow for the possibility of variable stellar H I line flux at the 5-20\% level, with a duty cycle of $\sim$3\%.  If observed in broad-band filters, these line flux enhancements would produce enhancements of 0.4 milli-mags (\textit{J}-band) to 0.3 milli-mags (\textit{Ks}-band), significantly below the upper limits of the IR photometric stability presented in  Section \ref{irflares}.

\section{Conclusions}

We have presented $\sim$47 hours of high cadence, high precision, simultaneous optical and IR photometric monitoring of 3 active M dwarfs.  We detected and characterized four highly energetic optical flares having U-band total energies of $\sim$7.8 x 10$^{30}$ to $\sim$1.3 x 10$^{32}$ ergs, and found no corresponding response in the \textit{J}, \textit{H}, or \textit{Ks} bandpasses at the precision of our data.  To summarize our results:

\begin{itemize}
\item For active dM3e stars, we find that a $\sim$1.3 x 10$^{32}$ erg \textit{U}-band flare will induce $<$8.3 (\textit{J}), $<$8.5 (\textit{H}), and $<$11.7 (\textit{Ks}) milli-mags of a response.  A flare of this energy or greater should occur less than once per 18 hours.
\item For active dM4.5e stars,  we find that a $\sim$5.1 x 10$^{31}$ erg \textit{U}-band flare will induce $<$7.8 (\textit{J}), $<$8.8 (\textit{H}), and $<$5.1 (\textit{Ks}) milli-mags of a response.  A flare of this energy or greater should occur less than once per 10 hours.
\item No evidence of stellar variability not associated with discrete flare events was observed at the level of $\sim$3.9 milli-mags over 1 hour time-scales and at the level of $\sim$5.6 milli-mags over 7.5 hour time-scales.  We therefore demonstrate that most M dwarf stellar activity and flares will not influence IR detection and characterization studies of M dwarf exoplanets above the level of $\sim$5-11 milli-mags, depending on the filter and spectral type.  
\item Based on the detection of the E$_{U}$ $>$10$^{34}$ ergs megaflare event \citep{kow10} in the (i+z) filter of the MEarth survey \citep{irw11}, and extrapolations of phenomenological flare models to IR (\textit{J,H,Ks}) wavelengths, we speculate that the most energetic megaflares on M dwarfs, which occur at rates of once per month, are likely to be easily detected in IR observations which obtain sensitivities of tens of milli-mags.  
\item Future IR transmission spectroscopic studies of M dwarf exoplanets should note that 5-20\% flux enhancements have been observed 
in H I Paschen and Br$\gamma$ lines during flares having total energies similar to event \#1 in our survey.  This level of line flux will only produce a small (0.3-0.4 milli-mag) enhancement in broad-band (\textit{J,H,Ks}) filter observations and should not preclude molecular line analysis.

\end{itemize}

\acknowledgements 

We thank the referee, Rachel Osten, for providing comments which helped to improve the content and clarity of this paper.  BMT acknowledges support from a Mary Gates Research Scholarship.  We acknowledge support from NSF AST grants 08-02230 (JPW),  08-07205 (AFK, ELH, SLH), and 06-45416 (PK).  AFK and SJS thank NOAO for supporting their travel to KPNO to carry out portions of these observations.  Observations from the NMSU 1m were supported in part by NSF AST 05-19398.  We thank J. Irwin for sharing MEarth detections of a YZ CMi flare with us, and J. Davenport for providing access to his model results and discussion on the topic.  BMT and JPW thank the Kailua-Kona shark, which was lurking in nearby waters during the preparation of this manuscript, for not eating them.

{\it Facilities:} \facility{ARC}, \facility{NMSU:1m}, \facility{KPNO:2.1m}, \facility{WIYN:0.9m}

\newpage
\clearpage
\begin{table}
\rotate
\begin{center}
\footnotesize
\caption{Summary of Observations\label{obssum}}
\begin{tabular}{lcccccc}
\tableline
UT Date & Science Object & Facility & Filter & Integration Time & Total Obs Time & Obs Cadence$^{1}$ \\
 & & & & sec & hours & \\
\tableline

2009 Oct 8 & EV Lac & WIYN 0.9m & U & 15 & 8 & 1 obs per 51 sec  \\
2009 Oct 8 & EV Lac & KPNO 2.1m & J,H,Ks & 0.87 & 8 & 1 obs per 52 sec  \\
2011 Feb 12 & YZ CMi & NMSU 1m & U & 10 & 8.25 & 1 obs per 18 sec$^{2}$  \\
2011 Feb 12 & YZ CMi & ARCSAT 0.5m & g & 1 &  8.25 & 2 obs over 4.2 sec; then 17 sec break  \\
2011 Feb 12 & YZ CMi & ARCSAT 0.5m & r & 0.5 &  8.25 & 3 obs over 4.8 sec; then 16 sec break  \\
2011 Feb 12 & YZ CMi & ARCSAT 0.5m & i & 0.5 &  8.25 & 4 obs over 6.0 sec; then 15 sec break  \\
2011 Feb 12 & YZ CMi & KPNO 2.1m & J,H,Ks & 0.87 & 8.25  & 1 obs per 60 sec \\
2011 Feb 12 & AD Leo & NMSU 1m & U & 4 & 3 & 1 obs per 12 sec$^{2}$ \\
2011 Feb 12 & AD Leo & ARCSAT 0.5m & g & 1 & 3 & 4 obs over 8.4 sec; then 27 sec break \\
2011 Feb 12 & AD Leo & ARCSAT 0.5m & r & 1 & 3 & 4 obs over 8.4 sec; then 27 sec break \\
2011 Feb 12 & AD Leo & ARCSAT 0.5m & i & 1 & 3 & 4 obs over 8.4 sec; then 27 sec break \\
2011 Feb 12 & AD Leo & KPNO 2.1m & J,H,Ks & 0.87 & 3 & 1 obs per 58 sec \\
2011 Feb 13 & YZ CMi & NMSU 1m & U & 10 & 7.5 & 1 obs per 18 sec$^{2}$ \\
2011 Feb 13 & YZ CMi & ARCSAT 0.5m & g & 1 & 7.5 & 4 obs over 8.4 sec; then 27 sec break \\
2011 Feb 13 & YZ CMi & ARCSAT 0.5m & r & 1 & 7.5 & 4 obs over 8.4 sec; then 27 sec break \\
2011 Feb 13 & YZ CMi & ARCSAT 0.5m & i & 1 & 7.5 & 4 obs over 8.4 sec; then 27 sec break \\
2011 Feb 13 & YZ CMi & KPNO 2.1m & J,H,Ks & 0.87 & 7.5 & 1 obs per 60 sec\\
2011 Feb 13 & AD Leo & NMSU 1m & U & 4 & 3 & 1 obs per 12 sec$^{2}$ \\
2011 Feb 13 & AD Leo & ARCSAT 0.5m & g & 1 & 3 & 4 obs over 8.4 sec; then 27 sec break \\
2011 Feb 13 & AD Leo & ARCSAT 0.5m & r & 1 & 3 & 4 obs over 8.4 sec; then 27 sec break \\
2011 Feb 13 & AD Leo & ARCSAT 0.5m & i & 1 & 3 & 4 obs over 8.4 sec; then 27 sec break \\
2011 Feb 13 & AD Leo & KPNO 2.1m & J,H,Ks & 0.87 & 3  & 1 obs per 58 sec \\
2011 Feb 15 & YZ CMi & NMSU 1m & U & 10 & 9 & 1 obs per 18 sec$^{2}$ \\
2011 Feb 15 & YZ CMi & ARCSAT 0.5m & g & 1 & 9  & 4 obs over 8.4 sec; then 27 sec break \\
2011 Feb 15 & YZ CMi & ARCSAT 0.5m & r & 1 & 9  & 4 obs over 8.4 sec; then 27 sec break \\
2011 Feb 15 & YZ CMi & ARCSAT 0.5m & i & 1 & 9  & 4 obs over 8.4 sec; then 27 sec break \\
2011 Feb 15 & YZ CMi & KPNO 2.1m & J,H,Ks & 0.87 - 1.200 & 9 & 1 obs per 58 sec \\
2011 Feb 16 & YZ CMi & NMSU 1m & U & 10 & 8 & 1 obs per 18 sec$^{2}$ \\
2011 Feb 16 & YZ CMi & ARCSAT 0.5m & g & 1 & 8  &  4 obs over 8.4 sec; then 27 sec break \\
2011 Feb 16 & YZ CMi & ARCSAT 0.5m & r & 1 & 8  &  4 obs over 8.4 sec; then 27 sec break \\
2011 Feb 16 & YZ CMi & ARCSAT 0.5m & i & 1 & 8  &  4 obs over 8.4 sec; then 27 sec break \\
2011 Feb 16 & YZ CMi & KPNO 2.1m & J,H,Ks & 0.87 & 8 & 1 obs per 58 sec \\

\tableline
\end{tabular}
\end{center}
\vspace{-0.3in}
\tablecomments{The integration times and effective observing cadence are tabulated for each of our observations.  Note that the observing cadence includes 
exposure times as well as read-out and data transfer overheads.  $^{1}$ All cadences compiled in column 7 are approximate.  $^{2}$ Indicates the most common observing cadence, although the robotic NMSU 1m telescope occasionally intersperses a single exposure separated by $\sim$24 seconds, and routinely pauses to refocus roughly every 60 exposures.}
\end{table}

\newpage
\clearpage
\begin{table}
\begin{center}
\footnotesize
\caption{Summary of Science and Comparison Stars\label{comp}}
\begin{tabular}{lccccc}
\tableline
Science Object & Spectral Type & H Magnitude & IR Comparison Star & Spectral Type & H magnitude \\
\tableline
YZ CMi & dM4.5e & 6.005 & HD 62525 & G5 & 6.035 \\
AD Leo & dM3e & 4.843 & TYC 1423-165-1 & G5 & 7.686 \\
EV Lac & dM3.5e & 5.554 & BD+43 4303 & K0V & 6.685 \\

\tableline
\end{tabular}
\end{center}
\vspace{-0.3in}
\tablecomments{Basic properties of our three science targets and the associated comparison 
stars used for computing differential photometry for each is compiled.}
\end{table}

\newpage
\clearpage
\begin{table}
\begin{center}
\footnotesize
\caption{Summary of Detected Flares\label{flaresum}}
\begin{tabular}{lccccc}
\tableline
UT Date & Star & Filter & \# Flares Detected & IR Stability (full night) & IR Stability (1 hr) \\
 & & & & milli-mag & milli-mag \\
\tableline
20091008 & EV Lac &  U	& 1	& \nodata & \nodata \\	
\nodata & \nodata & J & 0 &	9.0 & 5.7 \\
\nodata & \nodata & H & 0 & 8.9 & 4.5 \\
\nodata & \nodata & Ks & 0 & 8.7 & 6.5 \\
20110212 & YZ CMi & U & 6 & \nodata & \nodata \\
\nodata & \nodata & g & 1 & \nodata & \nodata \\	
\nodata & \nodata & r & 1 & \nodata & \nodata \\
\nodata & \nodata & i & 0 & \nodata & \nodata \\
\nodata & \nodata & J & 0 & 8.6	& 6.0 \\
\nodata & \nodata & H & 0 & 8.7 & 6.1 \\
\nodata & \nodata & Ks & 0 & 7.4 & 6.1 \\
20110212 & AD Leo & U & 0 & \nodata & \nodata \\
\nodata & \nodata & g & 0 & \nodata & \nodata \\
\nodata & \nodata & r & 0 & \nodata & \nodata \\
\nodata & \nodata & i & 0 & \nodata & \nodata \\
\nodata & \nodata & J & 0 & 10.0 & 9.6 \\
\nodata & \nodata & H & 0 & 9.7 & 8.0 \\
\nodata & \nodata & Ks & 0 & 12.6 &	10.5 \\
20110213 & YZ CMi & U & 4 & \nodata & \nodata \\
\nodata & \nodata & g & 0 & \nodata & \nodata \\
\nodata & \nodata & r & 0 & \nodata & \nodata \\
\nodata & \nodata & i & 0 & \nodata & \nodata \\
\nodata & \nodata & J & 0 & 6.0	& 3.9 \\
\nodata & \nodata & H & 0 & 5.6 & 3.8 \\
\nodata & \nodata & Ks & 0 & 5.1 & 3.9 \\
20110213 & AD Leo & U & 2 & \nodata & \nodata \\
\nodata & \nodata & g & 2 & \nodata & \nodata \\
\nodata & \nodata & r & 1 & \nodata & \nodata \\
\nodata & \nodata & i & 1 & \nodata & \nodata \\
\nodata & \nodata & J & 0 & 11.4 & 8.5 \\
\nodata & \nodata & H & 0 & 11.2 & 8.4 \\
\nodata & \nodata & Ks & 0 & 12.1 & 8.8 \\
20110215 & YZ CMi & U & 4 & \nodata & \nodata \\
\nodata & \nodata & g & 0 & \nodata & \nodata \\
\nodata & \nodata & r & 0 & \nodata & \nodata \\
\nodata & \nodata & i & 0 & \nodata & \nodata \\
\nodata & \nodata & J & 0 & 7.8	& 4.5 \\
\nodata & \nodata & H & 0 & 8.8	& 5.0 \\	
\nodata & \nodata & Ks & 0 & 8.3 & 4.5 \\
20110216 & YZ CMi  & U & 3 & \nodata & \nodata \\
\nodata & \nodata & g & 2 & \nodata & \nodata \\
\nodata & \nodata & r & 0 & \nodata & \nodata \\
\nodata & \nodata & i & 0 & \nodata & \nodata \\
\nodata & \nodata & J & 0 & 6.1 & 4.2 \\
\nodata & \nodata & H & 0 & 6.2 & 4.8 \\
\nodata & \nodata & Ks & 0 & 8.9 & 5.0 \\
\tableline
\end{tabular}
\end{center}
\vspace{-0.3in}
\tablecomments{The number of flares detected via the automated flare finding algorithm 
discussed in Section \ref{results} is summarized for each filter.  For the SQIID IR (JHKs) 
photometry, we also compile the observed standard deviation of the differential photometry 
across the entire night and in a 1-hour subset of each night, in units of milli-magnitudes.}
\end{table}

\newpage
\clearpage
\begin{table}
\begin{center}
\footnotesize
\caption{Summary of Flare Properties\label{flarenergy}}
\begin{tabular}{lcccccccc}
\tableline
UT Date & Star & Event \# & Filter & Peak $\Delta$ Magnitude  & Duration & Equiv Duration & Flare Energy  & Flare Frequency \\
 & & & & magnitude & sec. & sec. & ergs & hours$^{-1}$ \\
\tableline
20110213 & AD Leo & 1 & U	& 1.53 & 2460	& 797 &	$\sim$1.27e+32 & $\sim$0.056 \\
\nodata & \nodata & \nodata &	g & 0.25 &	1085 &	54 & \nodata & \nodata \\
\nodata & \nodata & \nodata & 	r & 0.10 & 	921 & 21 &	\nodata & \nodata  \\
\nodata & \nodata & \nodata &	i & 0.04 &	232 &	4 &	 \nodata & \nodata \\
\nodata & \nodata & \nodata &	J & $<$ 0.0083 & \nodata & \nodata & \nodata & \nodata \\
\nodata & \nodata & \nodata & H	& $<$ 0.0085	& \nodata & \nodata & \nodata & \nodata \\
\nodata & \nodata & \nodata & Ks & $<$ 0.0117 & \nodata & \nodata & \nodata & \nodata \\
20110212 & YZ CMi & 2 & U & 1.61 & 3574 & 1121 & $\sim$5.12e+31 & $\sim$0.100 \\
\nodata & \nodata & \nodata & g	& 0.30 & 621 & 52 &	\nodata & \nodata \\
\nodata & \nodata & \nodata & r	& 0.09 & 445 & 17 &	\nodata & \nodata \\
\nodata & \nodata & \nodata & i	& $<$ 0.0227 & \nodata & \nodata & \nodata & \nodata \\
\nodata & \nodata & \nodata & J	& $<$ 0.0078 & \nodata & \nodata & \nodata & \nodata \\
\nodata & \nodata & \nodata & H	& $<$ 0.0088 & \nodata & \nodata & \nodata & \nodata \\
\nodata & \nodata & \nodata & Ks & $<$ 0.0051 & \nodata & \nodata & \nodata & \nodata \\
20110213 & YZ CMi & 3 & U & 1.03 & 	3973 &	722 & $\sim$3.30e+31 & $\sim$0.128 \\		
\nodata & \nodata & \nodata & J	& $<$ 0.0067 & \nodata & \nodata & \nodata & \nodata \\
\nodata & \nodata & \nodata & H	& $<$ 0.0070 & \nodata & \nodata & \nodata & \nodata \\
\nodata & \nodata & \nodata & Ks & $<$ 0.0057 & \nodata & \nodata & \nodata & \nodata \\	
20091008 & EV Lac & 4 & U & 0.31 & 	904 & 118 &	$\sim$7.82e+30 & $\sim$0.142 \\
\nodata & \nodata & \nodata & J	& $<$ 0.0072	& \nodata & \nodata & \nodata & \nodata \\
\nodata & \nodata & \nodata & H	& $<$ 0.0102 & \nodata & \nodata & \nodata & \nodata \\	
\nodata & \nodata & \nodata & Ks & $<$ 0.0085 & \nodata & \nodata & \nodata & \nodata \\
\tableline
\end{tabular}
\end{center}
\vspace{-0.3in}
\tablecomments{The peak delta magnitude, duration, equivalent duration, energy, and flare frequency is given as a function of filter for each of 
our four flare events.  Note that positive peak delta magnitudes correspond to flux enhancements.  Section \ref{opticalflares} describes how flare equivalent durations, energies, and rates were computed.  All quoted peak magnitude upper limits represent the 1-$\sigma$ standard deviation of the observation data.}
\end{table}

\newpage
\clearpage
\begin{table}
\begin{center}
\footnotesize
\caption{Timing of Optical and IR Flares \label{flaretime}}
\begin{tabular}{lcccc}
\tableline
UT Date & Star & Event \# & Filter & Relative Time from Flare Peak  \\
 & & & & \\
\tableline
20110213 & AD Leo & 1 & U	& 0$\pm$10.4 sec\\
\nodata & \nodata & \nodata &	g & 21.5$\pm$1.6 sec after U-filter \\
\nodata & \nodata & \nodata & 	r & 2.4$\pm$1.6 sec after U-filter  \\
\nodata & \nodata & \nodata &	i & 9.7$_{-25.5}^{+1.6}$ sec after U-filter \\
\nodata & \nodata & \nodata &	J,H,Ks & 11 sec before U-filter; 46 sec after U-filter$^{1}$ \\
20110212 & YZ CMi & 2 & U & 0$\pm$13 sec \\
\nodata & \nodata & \nodata & g & 3.7$_{-15.8}^{+1.6}$sec after U-filter \\
\nodata & \nodata & \nodata & r	& 1.5 $_{-1.6}^{+15.8}$ sec after U-filter\\
\nodata & \nodata & \nodata & J,H,Ks	& 42 sec before U-filter; 18.5 sec after U-filter$^{1}$ \\
20110213 & YZ CMi & 3 & U & 0$\pm$13 sec \\		
\nodata & \nodata & \nodata & J,H,Ks	& 12 sec before U-filter; 48 sec after U-filter$^{1}$ \\
20091008 & EV Lac & 4 & U & 0 $\pm$ 43.5 sec \\
\nodata & \nodata & \nodata & J,H,Ks & 1.4 sec after U-filter$^{1}$ \\
\tableline
\end{tabular}
\end{center}
\vspace{-0.3in}
\tablecomments{We identified the relative epoch of peak flare emission in every filter that each of our four flare events were detected.  The quoted error bars 
on these time correspond to the time of the prior and subsequent integration.  $^{1}$We did not detect clear evidence of flares in the J,H,Ks filters; thus, for these filters we list the nearest 1-2 epochs of exposures relative to epoch of the peak $\Delta$ magnitude flux observed in the U-filter.}
\end{table}

\newpage
\clearpage

\newpage
\clearpage

\newpage
\clearpage

\newpage
\clearpage

\newpage
\clearpage

\newpage
\clearpage

\end{document}